\documentclass[slac_one]{revtex4}
\usepackage{graphicx}
\usepackage{epsfig}
\usepackage{fancyhdr}
\usepackage{wrapfig}
\newcommand{\GeV}{{\mathrm{GeV}}}
\newcommand{\pb}{{\mathrm{pb}}}
\newcommand{\nb}{{\mathrm{nb}}}
\newcommand{\us}{{\mathrm{\mu s}}}
\newcommand{\der}{{\mathrm{d}}}
\newcommand{\Oh}{{\mathcal{O}}}
\newcommand{\sub}[1]{_{\mathrm{#1}}}
\newcommand{\super}[1]{^{\mathrm{#1}}}
\newcommand{\artitle}[1]{}

\begin{document}

\title{Charm and Beauty Photoproduction at HERA} 

\author{B. List, representing the H1 and ZEUS Collaborations}
\affiliation{Institute for Experimental Physics, University of Hamburg,
D--22603 Hamburg, Germany}

\begin{abstract}
After the completion of data taking at HERA-2, a large data set
is at hand to study the photoproduction of charm and beauty 
quarks in $ep$ collisions.
New measurements of charm production based on $D^*$ meson tagging
and beauty production based on muon and electron reconstruction
test perturbative QCD calculations with improved accuracy.
In general, QCD calculations at NLO describe the data well.
The scale uncertainties are, however, large and dominate over the
experimental uncertainties, calling for more precise calculations.

\end{abstract}

\maketitle

\thispagestyle{fancy}

\section{INTRODUCTION}

The study of charm and beauty production is a central
topic of research at HERA. 
In $ep$ collisions,
charm and beauty quarks are predominantly produced by the photon gluon
fusion process, which makes the production rates directly sensitive to
the gluon density in the proton.
The mass of these heavy quarks provides a hard scale so that
calculations in perturbative QCD are expected to be reliable.

However, this seemingly simple picture of photon gluon fusion with a hard
scale given by the quark mass is considerably complicated by two issues: 
on one hand by the presence of additional hard scales such as the quark's transverse
momentum, which means that heavy flavour production at HERA is
inherently a multi-scale problem; on the other hand by the hadronic
structure of the photon that leads to additional contributions such as
flavour excitation, which pose additional challenges to theoretical
calculations.

While it is possible to take the stance that the theoretical issues are
sufficiently well under control that one can infer from the measured charm and
beauty production rates the gluon density of the
proton, a complementary approach is to use the gluon density as
determined from other processes, i.e. fits to inclusive structure
function data, in particular from HERA, and investigate the extent to
which the formation of heavy quarks is theoretically understood in a
r\'egime where multiple scales and QCD dynamics play an important r\^ole.  

In the following, new measurements of charm and beauty production in
photoproduction at HERA will be presented. 
Photoproduction is characterized by a near zero virtuality\footnote{The
experimental definition is usually $Q^2 < 1\,\GeV^2$ and is
driven by the detector acceptance for scattered electrons.}
$Q^2 \approx 0\,\GeV^2$ of the exchanged photon, which essentially behaves like
a real photon.

\section{THEORETICAL MODELS}

To correct experimental data for detector effects, Monte Carlo
generators based on leading order ($\Oh\,(\alpha\sub s)$) matrix elements
and matched parton showers are used.
The Pythia program \cite{bib:pythia} is based on collinear factorization
and DGLAP evolution of parton densities.
In contrast, Cascade \cite{bib:cascade} is based on the $k\sub T$ factorization
ansatz, which uses a $k\sub T$ unintegrated gluon density that is
evolved according to the CCFM evolution. 

For calculations at next to leading order, the FMNR program
\cite{bib:fmnr}, based on the calculation by Nason, Dawson and Ellis
\cite{bib:nde}, is available. In this program, which does not include parton
showers, the fragmentation to hadrons is usually modelled using an
independent fragmentation approach; a new development is the 
FMNR $\otimes$ {\scshape PYTHIA} program \cite{Geiser:2007py}, which
augments FMNR with full Lund string fragmentation.

\section{CHARM PRODUCTION}

\begin{figure}
\epsfig{file=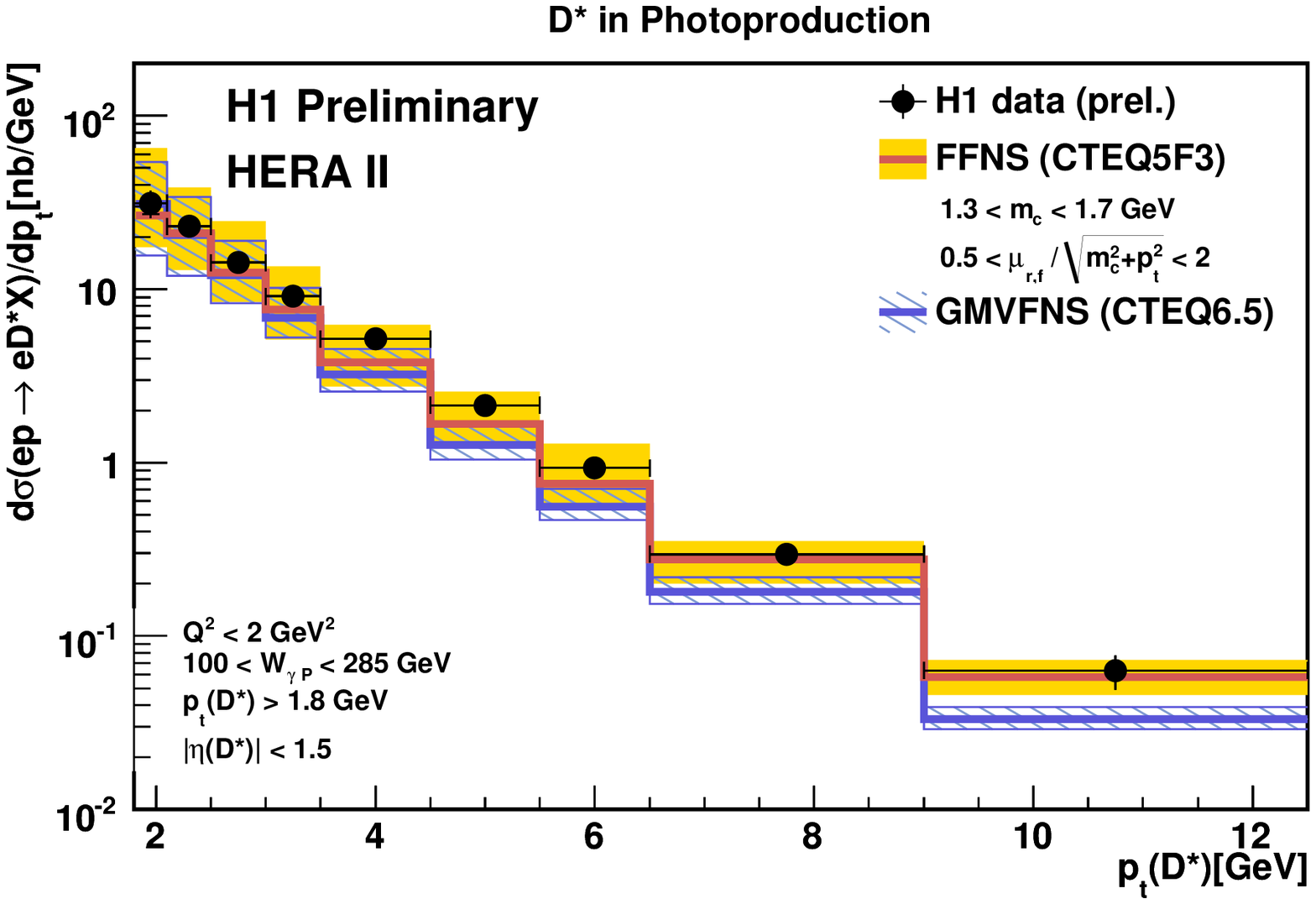,width=8.5cm}
\epsfig{file=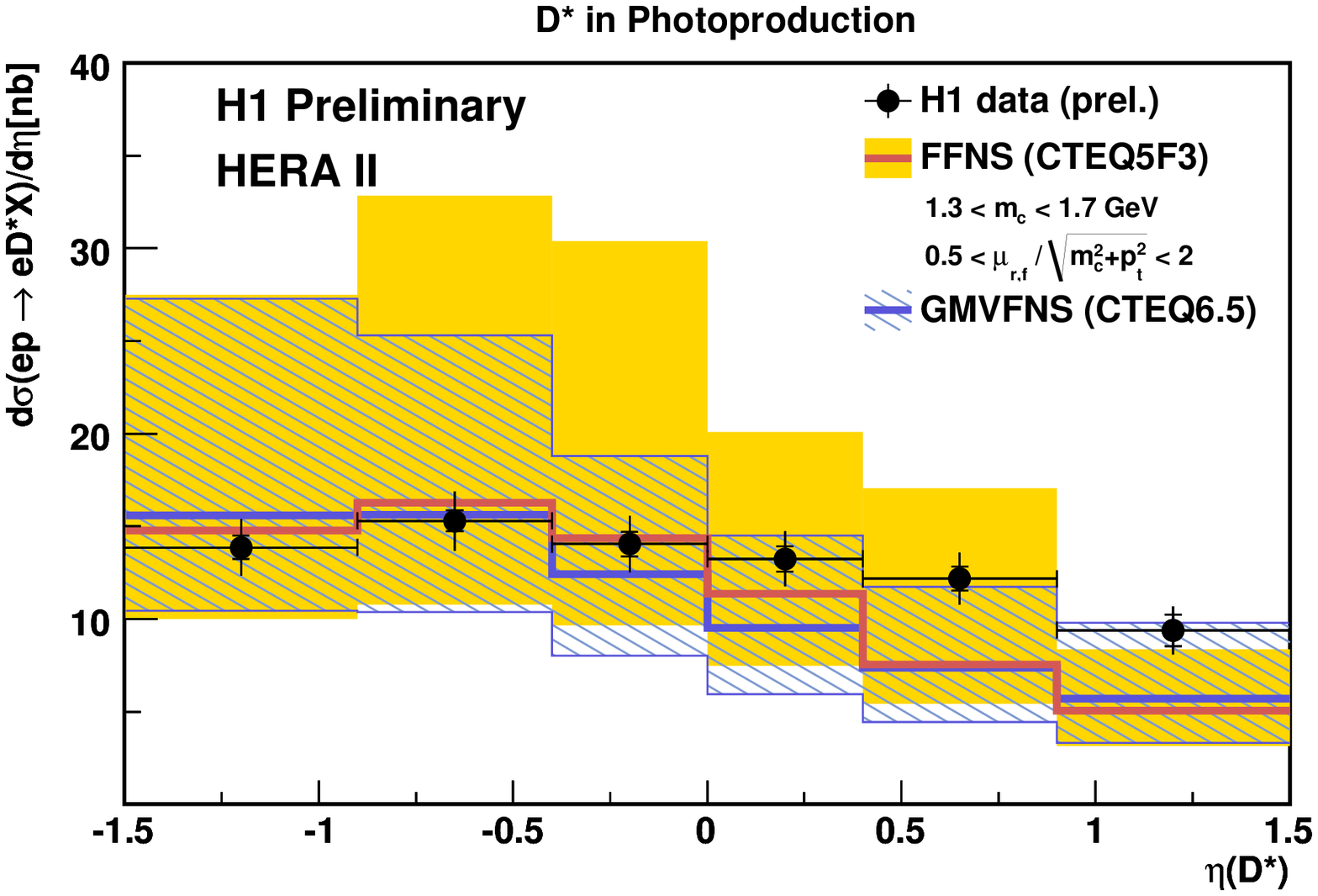,width=8.5cm}%
\caption{\label{fig:1} The H1 measurement of $D^*$ photoproduction,
compared to two QCD models at NLO: The FMNR program in the FFNS (shaded
area) and a new
calculation in the GM-VFNS (hatched). }
\end{figure}

The H1 collaboration has recently released a new measurement
\cite{bib:H1prelim-08-073}
 of charm
photoproduction based on an integrated luminosity of $93\,\pb^{-1}$ of
data taken at HERA-2.
Charm events were tagged by the presence of a $D^*$ meson decaying in
the golden decay channel $D^{*+} \to D^0 \pi^+$, $D^0 \to K^- \pi^+$
(and charge conjugates). The data were taken using a new trigger
\cite{Schoning:2004jp} that
allows a full mass reconstruction of the $D^*$ candidates at the third
trigger level within $100\,\us$.
The data covers a kinematic range defined by photon virtualities of
$Q^2 < 2\,\GeV^2$ and photon-proton
centre of mass energies of $100 < W\sub{\gamma p} < 285\,\GeV$.
The $D^*$ meson is measured for transverse momenta $p\sub T\,(D^*) >
1.8\,\GeV$ and pseudorapidities $|\eta\,(D^*)| < 1.5$.
Differential distributions in $p\sub T\,(D^*)$, $\eta\,(D^*)$ and
$W\sub{\gamma p}$ as well as double differential cross sections 
in  $p\sub T\,(D^*)$ and $\eta\,(D^*)$ have been measured and compared
to various QCD models.
In particular, two models at NLO QCD are available, namely FMNR and a
recent calculation in the GM-VFNS \cite{bib:gmvfns}.
Both models show similar behavour (Fig.~\ref{fig:1}): The $p\sub T\,(D^*)$ spectrum is
well described, with a slight deficiency at high $p\sub T$ in the case
of the GM-VFNS model, with sizeable uncertainties from scale variations,
in particular at low  $p\sub T$.
When the $\eta\,(D^*)$ spectrum is considered,
the theoretical uncertainties are several times larger than the
experimental ones; nevertheless, both models show a tendency to fall
below the data at large values of pseudorapidity. A similar behavour is
found in the deep-inelastic regime \cite{bib:krueger}. 

\section{BEAUTY PRODUCTION}

The measurement of beauty photoproduction at HERA is difficult due to
the fact that beauty production corresponds only to about $0.03\,\%$
of the total photoproduction cross section.
Two basic techniques have been used to enrich beauty production events:
The measurement of track impact parameters or secondary vertices, using
a silicon vertex detector, which enrich beauty production due to the
relatively large lifetime of beauty hadrons, and measurements based on
leptons (muons and/or electrons), which are based on the large
semileptonic branching fraction of beauty hadrons.

The H1 and ZEUS collaborations have recently released results from
several analyses based on lepton tags. 
Two analyses, one from H1 \cite{bib:H1prelim-08-071} and one from 
ZEUS \cite{bib:ZEUS-prel-07-020}, are based on samples with at least one muon and two jets in the
final state; in both cases, jet transverse momenta above $p\sub T \super
{jet1(2)} > 7 (6)\,\GeV$ are required. The resulting samples are
enriched to about $30\,\%$ of beauty events by the requirement of two
high-$p\sub T$ jets in conjunction with a highly energetic muon. The
actual fraction of beauty events is then determined from fits to the
distribution of the impact parameter $\delta$ of the muon track and the
relative transverse momentum $p\sub T \super {rel}$ of the muon with
respect to the nearest jet axis.
In these measurements, the $p\sub T \super {rel}$ distribution allows
the separation of beauty induced events from those of light (uds) and
charm quarks with good precision; the impact parameter distribution on
the other hand allows in addition the separation of light and charm 
quark induced events and thus serves not only as a cross check of the 
$p\sub T \super {rel}$ based measurement, but also helps to reduce the
systematic uncertainty which would otherwise be incurred by the
uncertainty of the charm quark contribution to the $p\sub T \super {rel}$
spectrum. Both measurements cover a range of $0.2 < y < 0.8$ in
inelasticity, momentum transfer $Q^2 < 1\,\GeV^2$, 
muon transverse momentum $p\sub T \super \mu > 2.5\,\GeV$
and jet pseudorapidity $|\eta \super {jet 1(2)}| < 2.5$;
however, the ZEUS data cover a range
$-1.6 < \eta\super{\mu} < 2.3$ of muon
pseudorapidities, while the H1 analysis is restricted to 
$-0.55 < \eta\super{\mu} < 1.1$.
The ZEUS analysis is based on $124\,\pb^{-1}$ of data taken in 2004 and
2005, while the H1 result uses $171\,\pb^{-1}$ of data from the years
2006 and 2007.

Both analyses show good agreement with perturbative QCD calculations
performed with the FMNR program, after corrections for hadronisation
effects have been applied to the FMNR predictions, which are available
only at parton level. The excess of data over NLO predictions 
at low values of jet and muon transverse momenta 
that was reported in an earlier H1 analysis of
HERA-1 data \cite{Aktas:2005zc} and was in disagreement with 
ZEUS data 
\cite{Chekanov:2003si}, is not confirmed by the new H1 analysis.

ZEUS has also published a new measurement based on events
with semileptonic decays to electrons \cite{Chekanov:2008aa}.
In this analysis, events with two jets of
$E\sub T \super {jet1 (2)} > 7 (6)\,\GeV$ and at least one electron are
selected from $120\,\pb^{-1}$ of data taken at HERA-1.
The beauty contribution was determined from a fit to a
likelihood variable based on quantities that are sensitive to the
quality of electron identification, such as the specific ionisation
$\der E/\der x$, as well as quantities that allow the separation between
electrons from light quark or charm induced processes from bottom
production. These quantities are again $p\sub T\super{rel}$, and 
$\Delta \phi$, the azimuthal angle difference between the 
electron flight direction and the direction of the total missing 
transverse momentum. Again, reasonable agreement with the NLO QCD
calculation FMNR is observed, for the total cross section as well as
differential distributions.

\begin{wrapfigure}{r}{0.5\columnwidth}
\epsfig{file=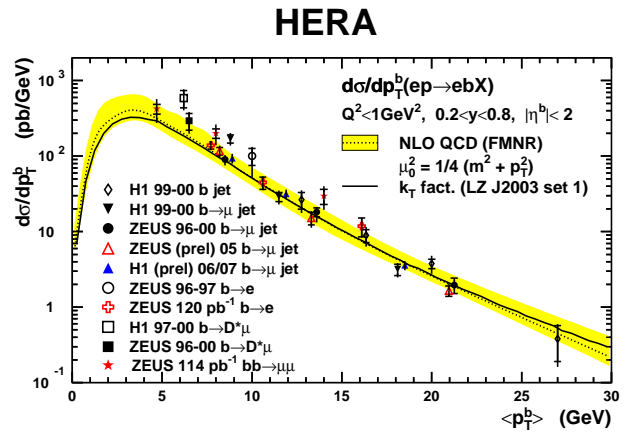,width=9.5cm}
\caption{\label{fig:2} A compilation of beauty photoproduction cross
section measurements as a function of the mean transverse momentum of
the beauty quarks, compared to QCD calculations at NLO. }
\end{wrapfigure}

An upcoming publication by ZEUS \cite{bib:zeus-dimuon} is based on
$114\,\pb^{-1}$ of data taken at HERA-1 and utilizes events with two
muons in the final state.
Contrary to other work discussed here, no jet requirement is imposed in
the selection, which is possible due to the high beauty fraction of a 
dilepton sample, and the analysis is performed down to muon transverse
momenta of $p\sub T^\mu > 1.5\,\GeV$, or even 
$p\sub T^\mu > 0.75\,\GeV$ for high-quality muon candidates. The
analysis is thus sensitive down to to the threshold of beauty
production, where the transverse momentum of the beauty quarks is close
to zero.
Therefore a measurement of the total beauty production
cross section is possible with relatively small extrapolation errors. 
The total cross section for the process 
$ep \to b \bar b X$ at $\sqrt s = 318\,\GeV$ 
has been determined to be 
$\sigma \sub {tot} (ep \to b \bar b X) = 13.9\,\pm
1.5\,({\mathrm{stat.}}) ^{+4.0}_{-4.3}\,({\mathrm{syst.}})\,\nb$,
to be compared to the NLO QCD prediction of
$\sigma \sub {tot}\super{NLO} (ep \to b \bar b X) = 7.5\,\pm
^{+4.5}_{-2.1}\,\nb$.
Within the large uncertainties, in particular of the NLO calculation,
the NLO prediction is consistent with the data.

Fig.~\ref{fig:2} summarizes all currently available beauty
photoproduction measurements at HERA, in comparison to the QCD
prediction at NLO. It can be seen that NLO QCD describes the data well
in shape and normalization, albeit with rather large uncertainties, which are
dominated by the choice of scale. Note that for this plot 
the central value of the QCD prediction is derived for a choice of
$\mu\sub {r, f} = \mu\sub {0} = 1 / 2 \sqrt{m\sub b^2 + p\sub T ^2}$
for the renormalization and factorization scales \cite{Geiser:2007tw},
while the individual measurements sometimes use 
twice this value, i.e.  
$\mu\sub {0} = \sqrt{m\sub b^2 + p\sub T ^2}$ for the central value.

\section{CONCLUSIONS}

The study of the photoproduction of charm and beauty quarks at HERA
remains a challenging testing ground for perturbative QCD.
With the increased statistical precision provided by the HERA-2 data
set, and improved detector understanding that results in more precise
measurements from HERA-1, measurements of charm and beauty production
are now available with experimental uncertainties that are significantly
smaller than the theoretical uncertainties of QCD calculations, even at
next to leading order.
Using gluon densities based on fits to inclusive structure function
measurements,
these calculations in general describe the data reasonably well in shape
and normalization, which is a great success of perturbative QCD. 
However, differential distributions such as the $\eta$ distribution of
$D^*$ mesons indicate deficiencies in the calculations which may need
further work to resolve.

\begin{acknowledgments}

This work was supported by the German Federal Ministry of Education and
Research.

\end{acknowledgments}

\end{document}